\documentclass[apj]{emulateapj-rtx4}
\usepackage{natbib}
\usepackage{amsmath}
\usepackage{amssymb}
\usepackage{amsthm}
\usepackage{bm}
\usepackage{dcolumn}
\usepackage{graphicx}
\usepackage{subfigure}
\usepackage{color}

\begin{document}

\title{Short-Lived $^{244}$Pu Points to Compact Binary Mergers as Sites for Heavy r-Process Nucleosynthesis}

\author{Kenta Hotokezaka}
\author{Tsvi Piran}
\author{Michael Paul}

\affiliation{Racah Institute of Physics, The Hebrew University, Jerusalem,
91904, Israel.}

\email{kenta.hotokezaka@mail.huji.ac.il}

\begin{abstract}
Measurements of the radioactive  $^{244}$Pu abundances
can break the degeneracy between high-rate/low-yield and low-rate/high-yield scenarios for the 
production of heavy $r$-process elements. The first corresponds to production by core collapse
supernovae~(cc-SNe) while the latter corresponds to production by e.g. compact binary mergers. 
The estimated $^{244}$Pu abundance in
the current interstellar medium inferred from deep-sea measurements~\citep{wallner2015NatCo}
is significantly lower than that corresponding  Early Solar
System abundances~\citep{turner2007E&PSL}.
We estimate the expected median value of the $^{244}$Pu abundances and fluctuations around this value 
in both models. We show that while the current and Early Solar System  abundances  
are naturally explained within the low-rate/high-yield (e.g. merger)
scenario, they are incompatible with the high-rate/low-yield~(cc-SNe) model.  
The inferred event rate remarkably agrees with compact binary merger rates estimated from Galactic neutron star
binaries
and from short gamma-ray bursts.
Furthermore, the ejected mass of $r$-process elements per event agrees with
both theoretical
and observational
macronova/kilonova estimates.
\end{abstract}
\maketitle

\section{Introduction}

The origin of heavy $r$-process elements is one of the current nucleosynthesis mysteries~
\citep{cowan1991PhR,qian2007PhR, arnould2007PhR}.
Core collapse supernovae \citep{burbidge1957RvMP}
and compact binary mergers are considered as possible sites~\citep{lattimer1974ApJ,eichler1989Nature,freiburghaus1999ApJ}.
The first produces small amounts of material at a high event rate while the latter produces large amounts in rare events.  
Radioactive elements, with the right lifetime can break the degeneracy between high-rate/low-yield and low-rate/high-yield scenarios.
Among radioactive elements most interesting is $^{244}$Pu (half-life of $81$~Million years) for which both the current
accumulation of live $^{244}$Pu particles accreted via interstellar particles in the Earth's deep sea floor~\citep{wallner2015NatCo}
and the Early Solar
System~(ESS) abundances have been measured~\citep{turner2007E&PSL}.
Interestingly, the estimated $^{244}$Pu abundance in
the current interstellar medium~(ISM) inferred from deep-sea measurements is significantly lower than that corresponding to the ESS measurements.

Using a model for the mixing of the heavy elements within the Galaxy we calculate the estimated median abundance and the fluctuation around this value 
in both scenarios and compare the results with astronomical observations. 

\section{Total heavy r-process production} 

\begin{figure*}
\includegraphics[bb = 0 0 210 210, width=100mm]{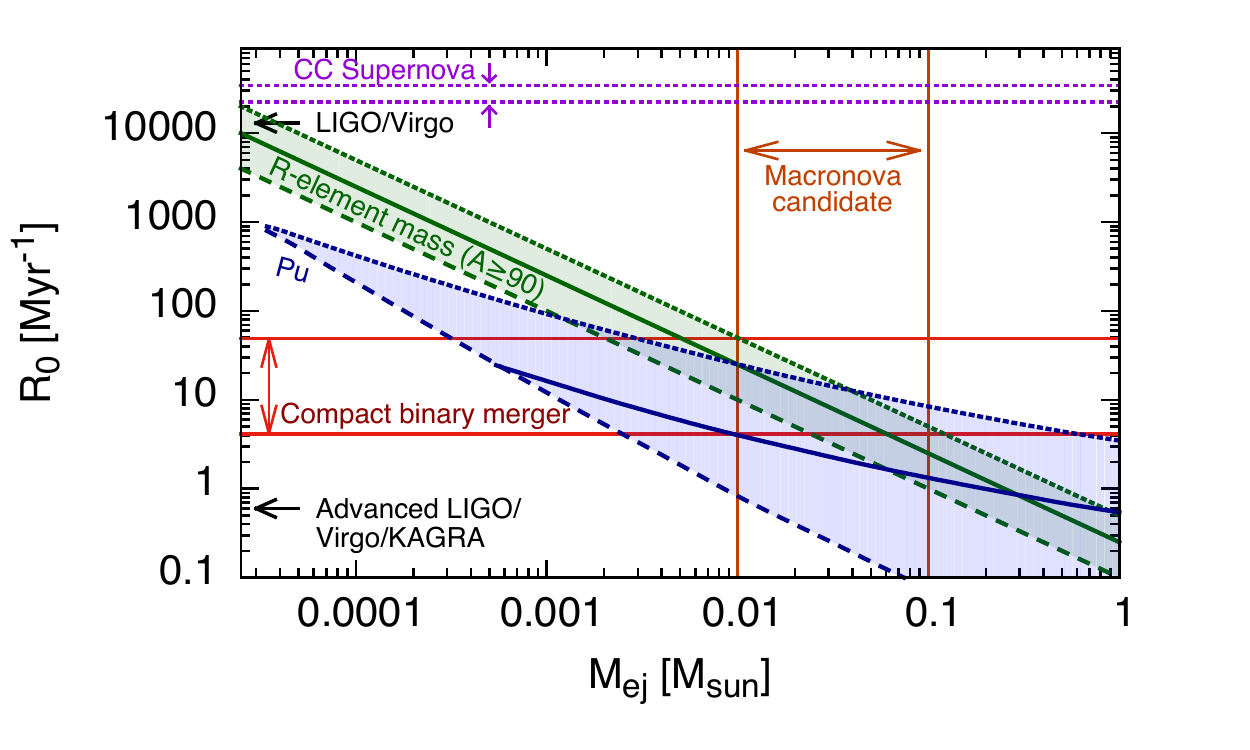}
\caption{\small{ The heavy $r$-process event rate and the ejected mass.
The diagonal green region expresses the degeneracy between high-rate/low-yield and low-rate/high-yield
corresponding to  the total mass of (stable)
$r$-process elements in the Galaxy, with
$R_{0}=\langle R\rangle$, $0.5\langle R\rangle$, and $0.2\langle R\rangle$~(see Eq.~\ref{R}).
The allowed region inferred from the $^{244}$Pu abundance in the deep-sea crust~\citep{wallner2015NatCo}
and the ESS~\citep{turner2007E&PSL, lodders2009}
is shown as a blue band. The  blue solid (dotted) line corresponds
to the  current ISM $^{244}$Pu density being the median ($2\sigma$) value.
The region above the dashed blue
line is the allowed region consistent  with the ESS measurement
(within $2\sigma$ fluctuations and
taking into account that the rate at $4.6$~Gyr BP can be
higher than $R_{0}$ by up to a factor of $5$).
The current event rate estimated from  binary neutron stars~\citep{kim2015MNRAS}
and  SGRBs~\citep{wanderman2015MNRAS}
are  shown
as the region between the horizontal red lines.
For SGRBs, we take an unknown jet beaming factor  in the
range  $10-70$. The region between the horizontal dotted purple lines corresponds to the cc-SNe  event rate
\citep{li2011MNRAS}.
Macronova mass estimates~\citep{rosswog2013RSPTA, hotokezaka2013PRDa,bauswein2013ApJa, tanvir2013Nature, berger2013ApJ,yang2015NatCo}
are between the vertical  dark orange lines.
The upper and lower horizontal arrows show the LIGO/Virgo upper limit of the merger~\citep{abadie2012PRD}
and the
expected capability of the
advanced gravitational-wave detectors with $5$~yr observations. The overlap region of the $^{244}$Pu
measurements and the total amount of heavy $r$-process element
s is consistent with that of the
compact binary merger scenario.}}
\label{fig:mr}
\end{figure*}

Using  the Solar abundance~\citep{goriely1999A&A}
as the mean value for stars
in the Galactic disk, the total mass of heavy $r$-process elements ($A\geq 90$) 
in the Galaxy is $M_{{\rm tot},A\geq 90}\approx 5\times 10^{3}M_{\odot}$.
These elements show a uniform abundance pattern
in metal-poor stars~\citep{sneden2008ARA},
suggesting that they are produced in a single kind of event.
The total mass  yields a relation between
the Galactic event rate,
$R$,  and the heavy $r$-process mass produced in each event $M_{{\rm ej},A\geq 90}$ (see Fig.~\ref{fig:mr}):
\begin{eqnarray}
\langle R\rangle \approx 50~{\rm Myr^{-1}}~\left(\frac{M_{{\rm ej},A\geq 90}}{0.01M_{\odot}} \right)^{-1} \ .  \label{R}
\end{eqnarray}
Here  $\langle R\rangle$  is the rate averaged over the age of the Galaxy
and it is not necessarily the same as the present-day event rate $R_{0}$.
For sources related to the death of massive stars the event rate should follow the star formation rate
which at present is lower than the average value~\citep{hopkins2006ApJ}.
For compact binary mergers, the event rate  follows with some delay the star formation rate.
The event rate of short gamma-ray bursts~(SGRBs; \citealt{wanderman2015MNRAS})
that likely arise from compact binary mergers~\citep{eichler1989Nature},
increases with the cosmological redshift $z$ at least up to $z\approx 0.8$.
In both cases, $R_{0}$ may be smaller than $\langle R \rangle$ by up to a factor of $\sim 5$.

\section{Measured short lived $^{244}$Pu abundances}

Using the total mass alone we cannot distinguish between the high-rate/low-yield  and the low-rate/high-yield sources.
Measurements  of abundances of short-lived radioactive $r$-process nuclides
can, however,  remove this degeneracy, as
these abundances  reflect the $r$-process production history on
timescales comparable to their lifetimes (modulo the Galactic mixing timescale). Among the various radioactive nuclides, $^{244}$Pu seems most suitable:
(i) it is produced only via the $r$-process;
(ii) the half-life of $^{244}$Pu, $81~$Million years~(Myr), is sufficiently short compared to
the Hubble time, while long enough to allow for significant accumulation;
(iii) both current and ESS~($\sim 4.6~{\rm Gyr}$ before present; BP)
$^{244}$Pu abundance in the ISM have been  measured.

The inner Solar System continuously accretes interstellar dust grains~\citep{mann2010ARA}
containing recently produced live radioactive $^{244}$Pu.
\citep{wallner2015NatCo}
(see also \citealt{paul2001ApJ})
measured the accumulation of $^{244}$Pu in a deep-sea
crust sample during the last $25$~Myr and estimated the  $^{244}$Pu flux on the Earth's orbit
as $250^{+590}_{-205}~{\rm cm^{-2}~Myr^{-1}}$,
where the upper and lower values correspond to $2\sigma$ limits\footnote{\cite{wallner2015NatCo}
analyze two samples, crust and sediment.
Here we focus on the crust sample that statistically dominates over the sediment one and spans a longer accumulation time (see Appendix). } .
The corresponding mean number density of $^{244}$Pu in the ISM is
$6\times 10^{-17}~{\rm cm^{-3}}$ and the $2\sigma$ upper limit is
$2\times 10^{-16}~{\rm cm^{-3}}$~(see Appendix).
These values are significantly lower than the number density in the ISM determined from the ESS $^{244}${\rm Pu} abundance:
$n_{_{\rm Pu}}\approx (^{244}{\rm Pu}/^{238}{\rm U})_{_{\rm ESS}}Y_{_{\rm U,\,ESS}}n_{_{\rm ISM}}\sim 6\times 10^{-15}~{\rm cm^{-3}}$.
The relative abundance ratio of $(^{244}{\rm Pu}/^{238}{\rm U})_{_{\rm ESS}}\approx 0.008$
is estimated from fissiogenic xenon in ESS material~\citep{turner2007E&PSL}.
 $Y_{_{\rm U,\,ESS}}=7.3\times 10^{-13}$ is
the number abundance of $^{238}$U relative to hydrogen inferred from meteorites~(\citealt{lodders2009} 
; corrected from the present for  $^{238}$U decay)
and $n_{_{\rm ISM}}$ is the mean number density of the ISM, which is typically $\approx 1$ cm$^{-3}$.

\section{Chemical mixing processes and median $^{244}$Pu abundance}

The abundance of a radioactive nuclide at a given location around the solar circle, $r_{\odot}$,
depends on the event rate density,
${\mathcal R}\equiv\rho_{*}(r_{\odot},0)R/M_{*}\approx 0.0015R~{\rm Myr^{-1}~kpc^{-3}}$, where
$\rho_{*}(r,z)$ is the stellar mass density in the disk,
$r$ and $z$ are the Galactic radius and height above the Galactic plane,
and $M_{*}$ is the total stellar mass in the disk~\citep{mcMillan2011MNRAS}.

The abundance depends also on the mixing timescale.
Heavy nuclei ejected into the ISM are homogenized via different processes
on different timescales~\citep{roy1995A&A}.
Initially, the gas containing heavy nuclei expands into the ISM as a blast wave until after  a couple of Myr~\citep{cioffi1988ApJ},
depending on the ejecta's kinetic energy, initial velocity and external density.
In the literature, it has been often assumed that at this stage the ejected material
remains within this fluid element, yielding a very slow process of
mixing within the galaxy. However,
the ISM turbulence  efficiently homogenizes
the ejected material~\citep{scalo2004ARA}.
The diffusion timescale
of the nuclei is determined by the turnover timescale of the largest eddies~\citep{pan2010ApJ}.
The diffusion coefficient can be described as
$D\equiv v_{\rm t}l_{\rm mix}/3\simeq \alpha~{\rm kpc^2/Gyr}~(v_{\rm t}/7~{\rm km~s^{-1}})
(H/0.2~{\rm kpc})$, where $v_{t}$ is the typical turbulent velocity of the
ISM, $l_{\rm mix}=3\alpha H$ is the turbulence mixing length
and $H$ is the scale-height of the local ISM.
Here we have introduced  a mixing length parameter $\alpha$, choosing $\alpha = 0.1$ as a reference value.
A numerical simulation of the turbulent mixing in the Galactic disk
shows this level of  efficiency of the mixing~\citep{yang2012ApJ}.
Defining  $\tau_{\rm mix}$ as the mean time between injection events at a given location,
$ ( 4\pi{\mathcal R} \tau_{\rm mix}/3 )^{-1/3} \equiv  2( D \tau_{\rm mix})^{1/2}$,
we have:
\begin{eqnarray}
\tau_{\rm mix} & \approx  300~{\rm Myr}~(R/10~{\rm Myr})^{-2/5}(\alpha/0.1)^{-3/5}\nonumber \\
 & (v_{t}/7~{\rm km/s})^{-3/5}(H/0.2~{\rm kpc})^{-3/5}.
\label{eq.2}
\end{eqnarray}

The median number density\footnote{The median rather than the average reflects
the density that a typical observer
measures.}   of a short-lived radioactive nuclide with
a mean-life $\tau_{i}$ is:
\begin{eqnarray}
\langle n_{i}\rangle_{m} \approx
n_{{\rm eq},i}\exp \left(-\frac{\tau_{\rm mix}}{2\tau_{i}}\right),
\end{eqnarray}
where $n_{{\rm eq},i}\approx N_{i}{\mathcal R}\tau_{i}$ is the equilibrium value and
$N_{i}$ is the total number of the nuclide $i$ ejected by each event.

For $\tau_{i}\gg \tau_{\rm mix}$, a typical observer measures $n_{{\rm eq},i}$.
For $\tau_{i}\ll \tau_{\rm mix}$, a typical observer measures
a number density much lower than $n_{{\rm eq},i}$ and one needs a larger  yield
to reach an observed value, interpreted here as the  median number density.
Figure~\ref{fig:mr} depicts the needed rate and yield so that the current
$^{244}$Pu is the median value for typical values of $\alpha$, $v_t$ and $H$ (see Eq. \ref{eq.2}). 
This relation (blue area in Fig.~\ref{fig:mr}) becomes flatter
than $R\propto M_{\rm ej}^{-1}$~(Eq.~\ref{R}, green band in Fig.~\ref{fig:mr})
for decreasing event rates breaking the rate-yield degeneracy.

\section{Monte-Carlo simulation of $^{244}$Pu abundance}

In order to take into account the large fluctuation in the  measured  number density averaged  over timescales
shorter than $\tau_{\rm mix}$,
we simulate the history of the $^{244}$Pu abundance in the ISM
around the solar circle over the last $7$~Gyr. We take into account the radioactive decay,
the turbulent diffusion process and the  time evolution of the production rate.
The $r$-process events are generated
randomly in a $4$-dimensional box with dimensions $7~{\rm Gyr}$ in time,
$16.66\pi~{\rm kpc}$ in the $x$-direction~(the circumference of
the solar circle), $2~{\rm kpc}$ in the $y$-direction~(the width of the circle),
and an exponential decay in the $z$-direction~(the height from the Galactic plane).
The events are distributed following
the stellar mass distribution in the Galactic disk~\cite{mcMillan2011MNRAS}
and the redshift
evolution following the SGRB rate~\citep{wanderman2015MNRAS}
or the cosmic star formation
history~\citep{hopkins2006ApJ}.
Each event ejects a fixed amount of $r$-process material
heavier than $A=90$ with the solar abundance pattern~\citep{goriely1999A&A, lodders2009}.
The number density of a radioactive nuclide
at a given time $t$ and a point $\vec{r}$
is computed by
\begin{eqnarray}
n_{i}(t,\vec{r}) = \sum_{j\in t>t_{j}} \frac{N_{i}}{K_{j}(t)}
\exp\left(-\frac{|\vec{r}-\vec{r}_{j}|^{2}}{4D\Delta t_{j}}-\frac{\Delta t_{j}}{\tau_{i}}\right),
\end{eqnarray}
where $\vec{r}_{j}$ and $t_{j}$ are the location and time of an event labeled by $j$,
$\Delta t_{j}\equiv t-t_{j}$, and
\begin{eqnarray}
K_{j}(t) =
{\rm min}
\left\{
\left(4\pi D \Delta t_{j} \right)^{3/2},~8\pi HD \Delta t_{j}
 \right\}
\end{eqnarray}
where the density evolution changes from the $3$-dimensional evolution to
the $2$-dimensional one appropriately.
For $^{244}$Pu particles, the total number is given by
$N_{\rm Pu}=(^{244}{\rm Pu}/^{238}{\rm U})_{0}N_{_{\rm ^{238}U}}$,
where $N_{_{\rm ^{238}U}}$ is the number of $^{238}$U particles and
$(^{244}$P/$^{238}$U$)_{0}$ is the initial production ratio.
This quantity depends on the details of the ejecta properties as well as on the
nuclear fission model. In the context of cc-SNe,
\cite{cowan1987ApJ}
estimated that a ratio of $0.4$
reproduces the solar abundance pattern.
For compact binary merger ejecta, \cite{eichler2015ApJ}
computed a production ratio of $0.33$. Here we use $(^{244}$Pu/$^{238}$U$)_{0}=0.4$.

\begin{figure*}
\includegraphics[bb = 0 0 350 260, width=170mm]{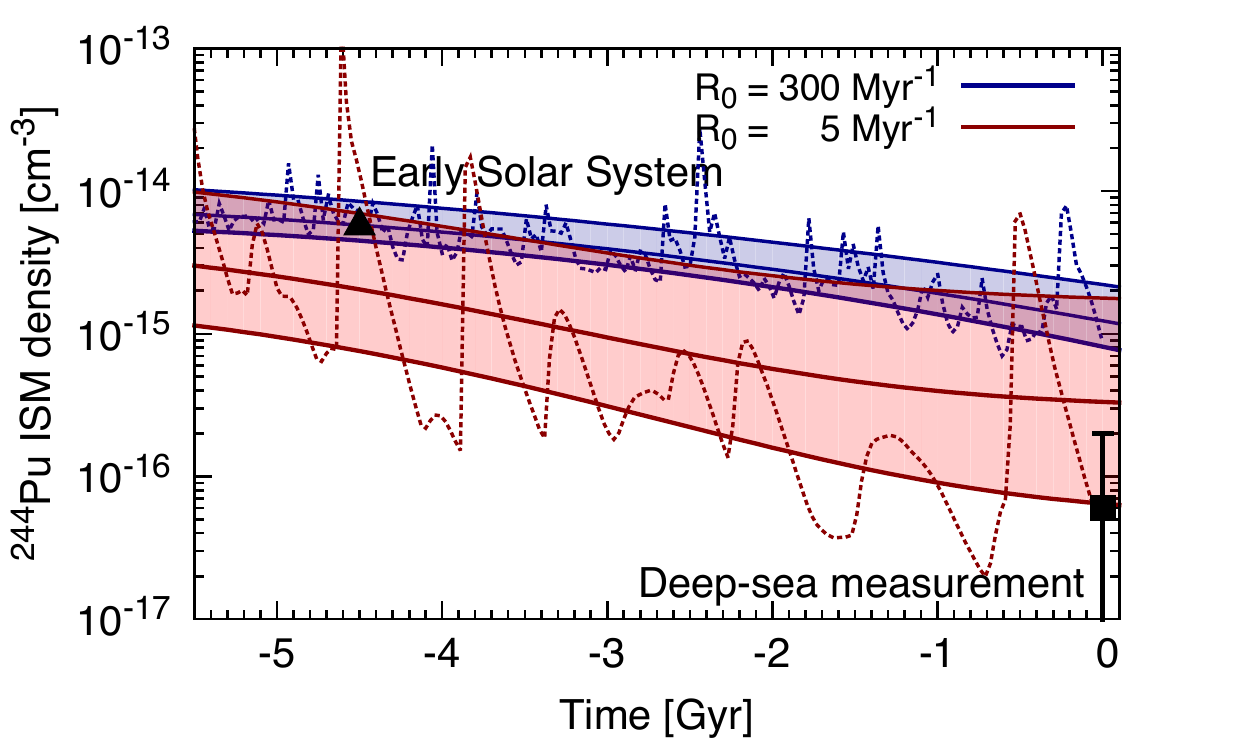}
\caption{Time evolution of $^{244}$Pu number densities in the ISM on the solar circle.
The solid red~(blue) lines represent the median number density and $\pm 1\sigma$
fluctuations for $R_{0}=5~{\rm Myr^{-1}}$~($R_{0}=300~{\rm Myr^{-1}}$).
The lower square with an error bar
shows $^{244}$Pu density with $2\sigma$ limits inferred from the deep sea measurement~\citep{wallner2015NatCo}.
The triangle at $4.6$~Gyr BP shows the value
at the time of the ESS~\citep{turner2007E&PSL, lodders2009}.
The production rate of $^{244}$Pu follows the time evolution of the SGRB
rate~\citep{wanderman2015MNRAS}
for $R_{0}=5~{\rm Myr^{-1}}$
and the cosmic star formation history~\citep{hopkins2006ApJ}
for $R_{0}=300~{\rm Myr^{-1}}$.
Also shown is an example of  the time sequence of $^{244}$Pu number densities at a given location
on the solar circle from a Monte-Carlo simulation for $R_{0}=5~{\rm Myr^{-1}}$~(dotted red line)
and $R_{0}=300~{\rm Myr^{-1}}$~(dotted blue line).
}
\label{fig:time}
\end{figure*}

We consider here a characteristic low-rate/high-yield case of $R_{0}=5~{\rm Myr^{-1}}$
following the SGRB rate evolution~\citep{wanderman2015MNRAS}
and
a high-rate/low-yield case of $R_{0}=300~{\rm Myr^{-1}}$ following the
cosmic star formation history~\citep{hopkins2006ApJ}.
Figure~\ref{fig:time} shows the results.
As expected  the fluctuations of  the low-rate/high-yield case are
much larger than those  of  the high-rate/low-yield one.
For both cases, the estimated range of number densities around $4.6$~Gyr BP are consistent with
the ESS values and they decrease with time
following the  decreasing  event rate. While for $R_{0}=5~{\rm Myr^{-1}}$
the simulated values are also consistent with the current deep-sea measurements,
for $R_{0}=300~{\rm Myr^{-1}}$ the  decline is insufficient  even when taking the fluctuations into account.

Figure~\ref{fig:mr} depicts  upper and lower bounds on the event
rate  which  consistently explain the $^{244}$Pu abundance of the ESS and the current ISM.
The sources must satisfy $R_{0}\leq 90~{\rm Myr^{-1}}$ and $ M_{\rm ej} \geq 0.001~M_{\odot}$.
While these limits vary somewhat with different assumed parameters,
 the qualitative result that we reach is robust
and independent of these choices. We conclude that unless some unknown process suppresses the present amount of $^{244}$Pu that reaches Earth, 
the heavy r-process sources are dominantly  low-rate/high-yield ones.

The results should be compared  with astronomical observations concerning the possible sources.
The low rate clearly rules out  cc-SNe. The current $^{244}$Pu abundance should be larger by  a factor of $5$
to $100$ to be compatible with a dominant cc-SNe source.
Turning  to compact binary mergers  Fig.~\ref{fig:mr} depicts also  (i) the merger rate estimated from known Galactic binary 
neutron stars~\citep{kim2015MNRAS}
and from the current SGRB rate~\citep{wanderman2015MNRAS}
and  (ii) the ejected mass of $r$-process elements estimated from macronova candidates associated with 
GRB~130603B~\citep{tanvir2013Nature, berger2013ApJ}
and with GRB~060614~\cite{yang2015NatCo}
and theoretical ejecta mass estimates~\citep{rosswog2013RSPTA, hotokezaka2013PRDa, bauswein2013ApJa}.
Remarkably, the rates and masses estimated here are fully consistent with those  observations.
In fact most of the overlap between the allowed $^{244}$Pu region and the overall $r$-process
production range is just in this part of the astrophysical parameter phase space describing
compact binary mergers  and macronova ejection estimates.

\section{Sensitivity of the results to the choice of parameters and redshift evolution}\label{sec:sen}
\begin{figure*}[h]
\includegraphics[bb = 0 0 380 210, width=160mm]{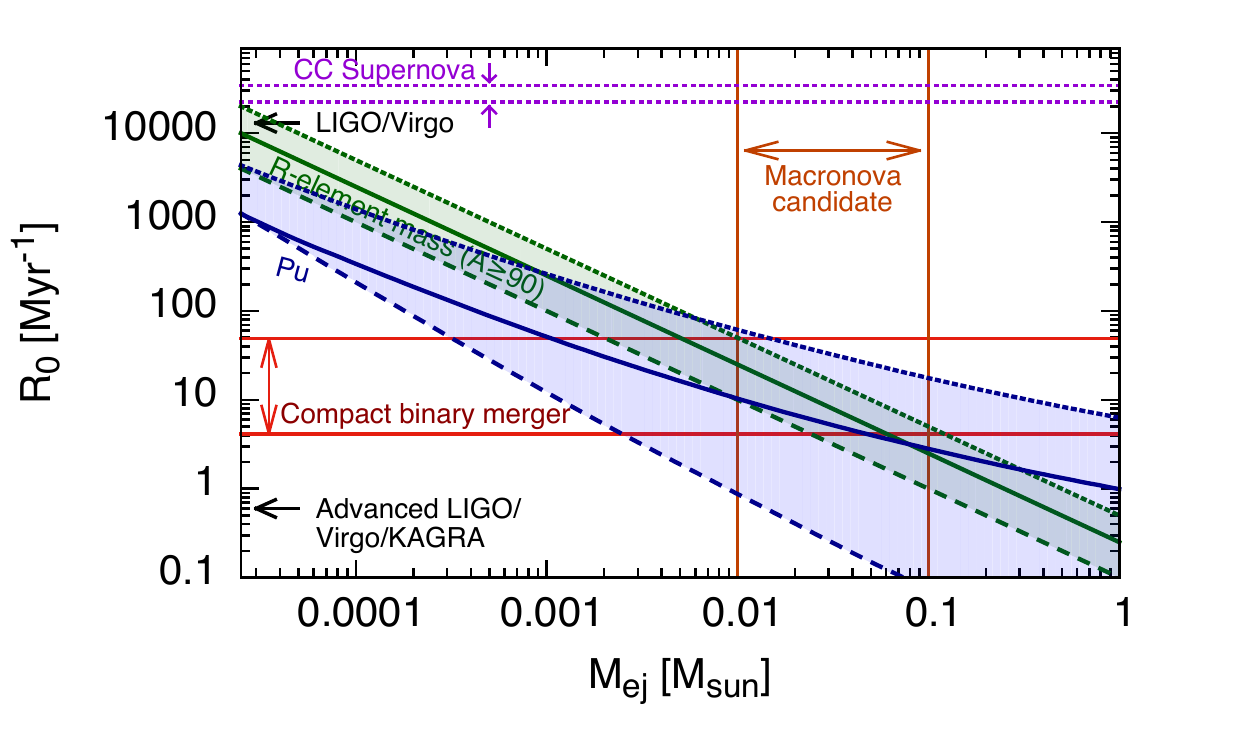}\\
\vspace{0.5cm}
\includegraphics[bb = 0 0 380 210, width=160mm]{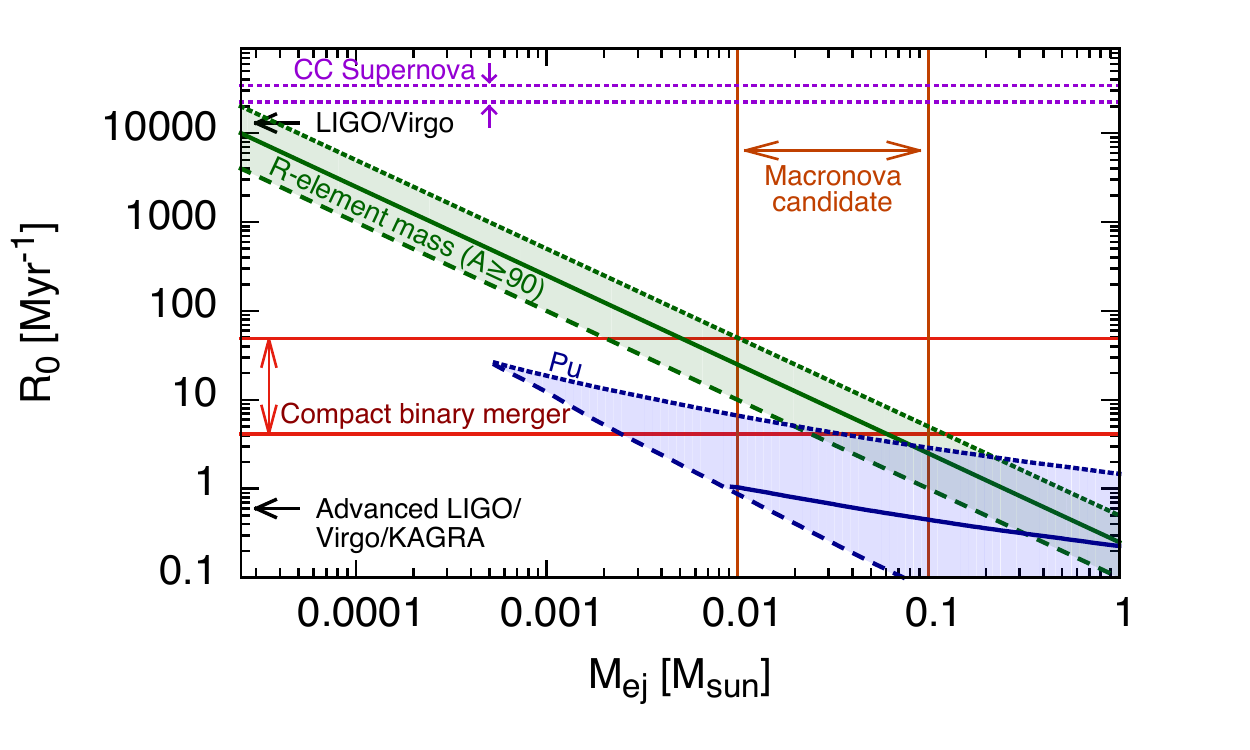}
\caption{
Same as Fig.~1 
but for the efficiency parameter $\epsilon=0.01$~(top)
and $0.9$~(bottom).
}
\label{fig:epsilon}
\end{figure*}

Before proceeding to the conclusion, we show the sensitivity
of the results to the choice of parameters.
Estimates of the rates  and yields
involve two unknown parameters: a parameter $\epsilon$ and the
mixing-length parameter $\alpha$~(see Appendix for details of $\epsilon$).
The first, $\epsilon$, introduces the largest uncertainty in the results.
For larger values of $\epsilon$,  smaller yields are sufficient,
as the depletion of the current ISM $^{244}$Pu is more significant.
Figure~\ref{fig:epsilon}
shows the rate-yield estimates
for $\epsilon=0.01$~(top panel) and for $0.9$~(bottom panel).
For $\epsilon = 0.01$, the estimated rate is $R_{0}<7000~{\rm Myr^{-1}}$ within the
$2\sigma$ level. Even though this rate is quite high  it is still  smaller than the rate
of normal cc-SNe  by a factor of a few.
For $\epsilon=0.9$, the allowed event rate is small $R_{0}<7~{\rm Myr^{-1}}$.

\begin{figure*}[h]
\includegraphics[bb = 0 0 380 210, width=160mm]{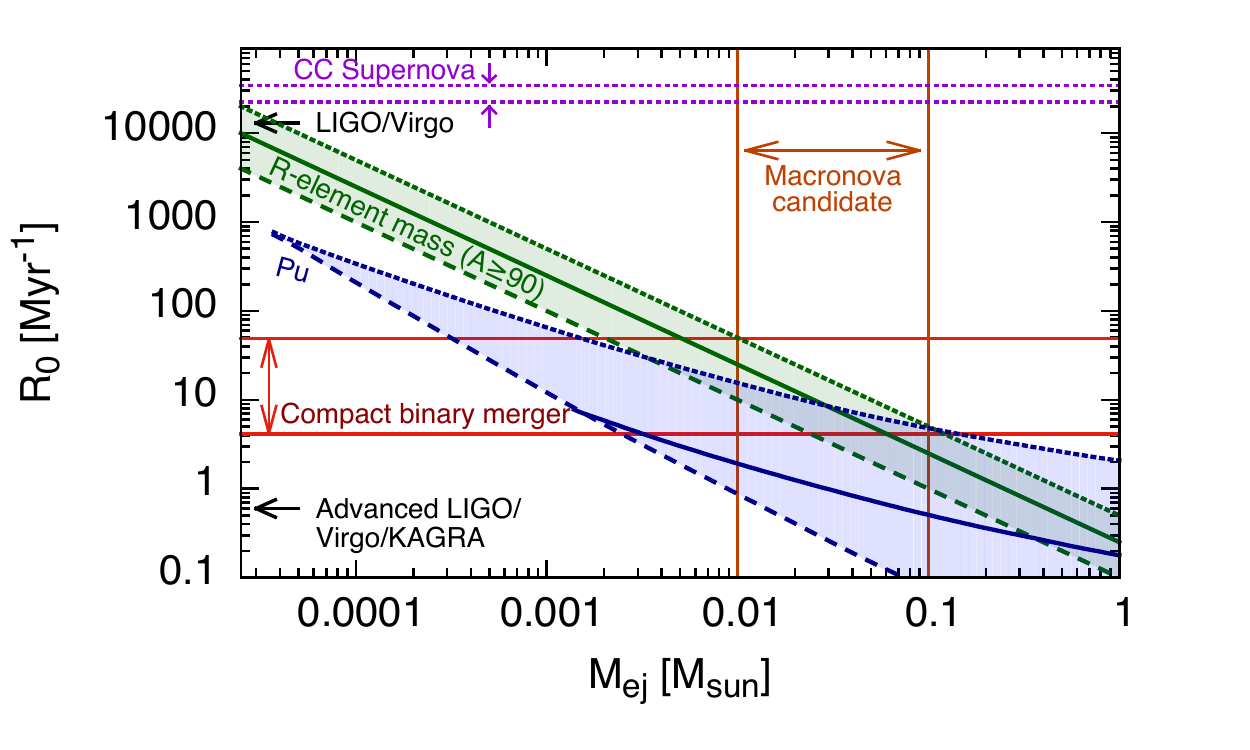}\\
\vspace{0.5cm}
\includegraphics[bb = 0 0 380 210, width=160mm]{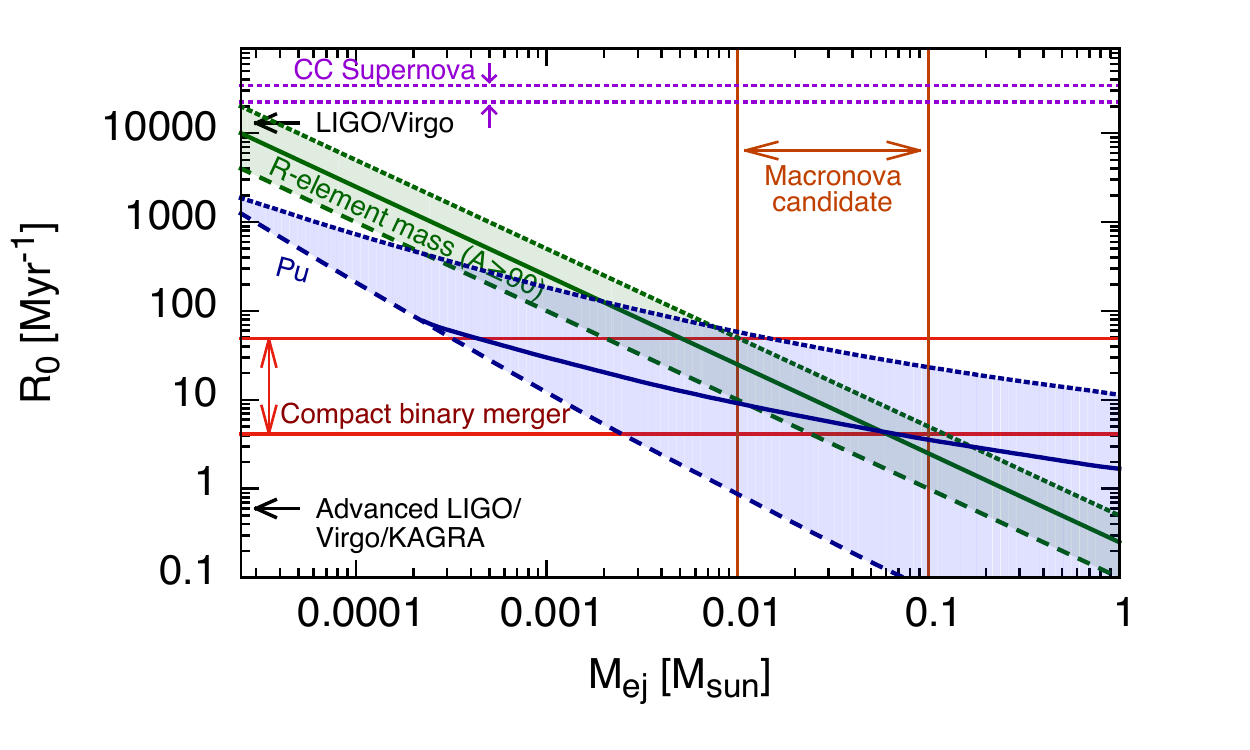}
\caption{
Same as Fig.~1 
but for the mixing parameter $\alpha=0.3$~(top) and $\alpha=0.03$~(bottom).
}
\label{fig:alpha1}
\end{figure*}

\begin{figure*}
\includegraphics[bb = 0 0 380 210, width=160mm]{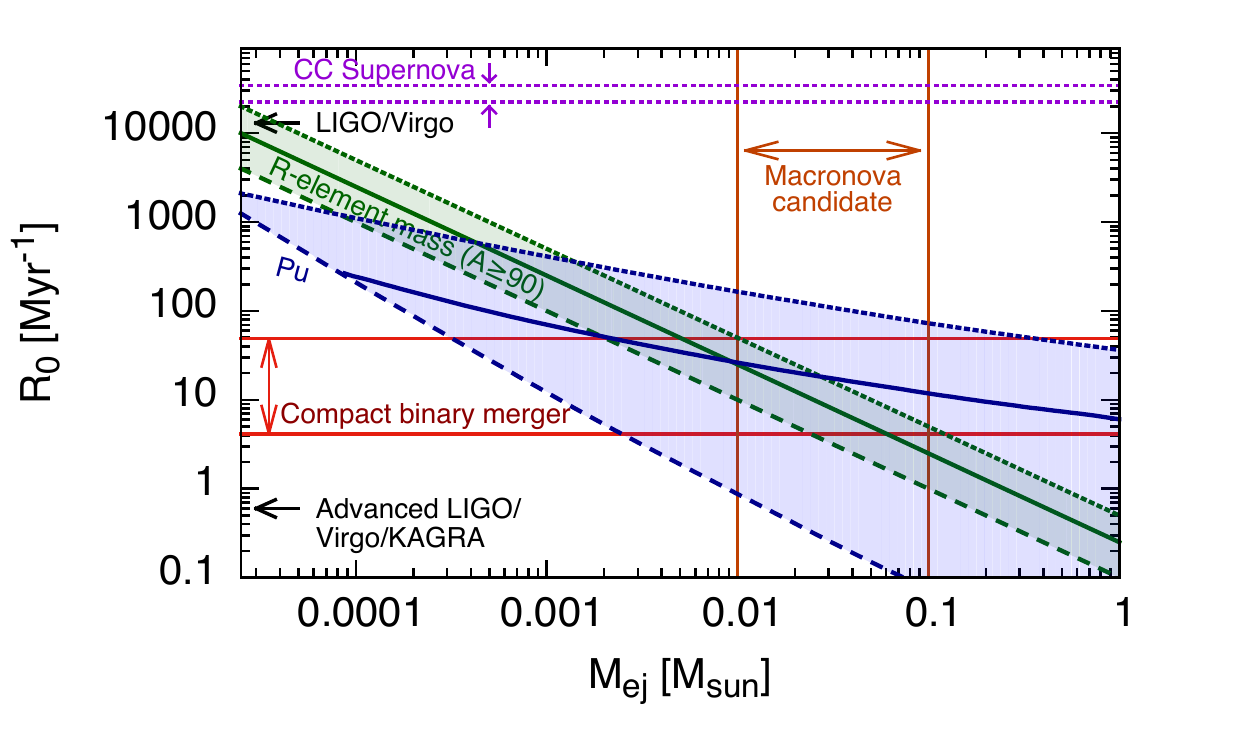}
\caption{
Same as Fig.~1 but for the mixing parameter $\alpha=0.01$.\hfill
}
\label{fig:alpha2}
\end{figure*}

The rate-yield  estimates with  different choices of $\alpha$~($0.3$,
$0.03$, and $0.01$) are shown in Fig.~\ref{fig:alpha1} and Fig.~\ref{fig:alpha2}.
For $\alpha=0.3$~(top panel of Fig.~\ref{fig:alpha1}), 
the mixing timescale
is shorter, implying that a single event can injects live $^{244}$Pu
particles into a larger volume. As a result, observes measure larger $^{244}$Pu
densities and the allowed rate-yield region in the figure shifts to the lower event rate
compared to those with $\alpha=0.1$.
On the contrary, with a smaller,  $\alpha=0.03$~(bottom panel of Fig.~\ref{fig:alpha1})
and $0.01$ in Fig.~\ref{fig:alpha2},
higher rates and smaller yields are allowed.
Although the overlap region of the $^{244}$Pu measurements and total mass of $r$-process elements
depends on the value of $\epsilon$ and $\alpha$, the estimate ranges of the rate and
yield are consistent with those of compact binary mergers irrespective
of the exact choice of these two parameters.
Thus we can generally conclude that the low-rate/high-yield scenario, or more specifically the
compact binary merger scenario  is preferred while the high-rate/low-yield scenario, or more specifically the cc-SNe scenario,
is ruled out.

\begin{figure*}[h]
\includegraphics[bb = 0 0 380 210, width=160mm]{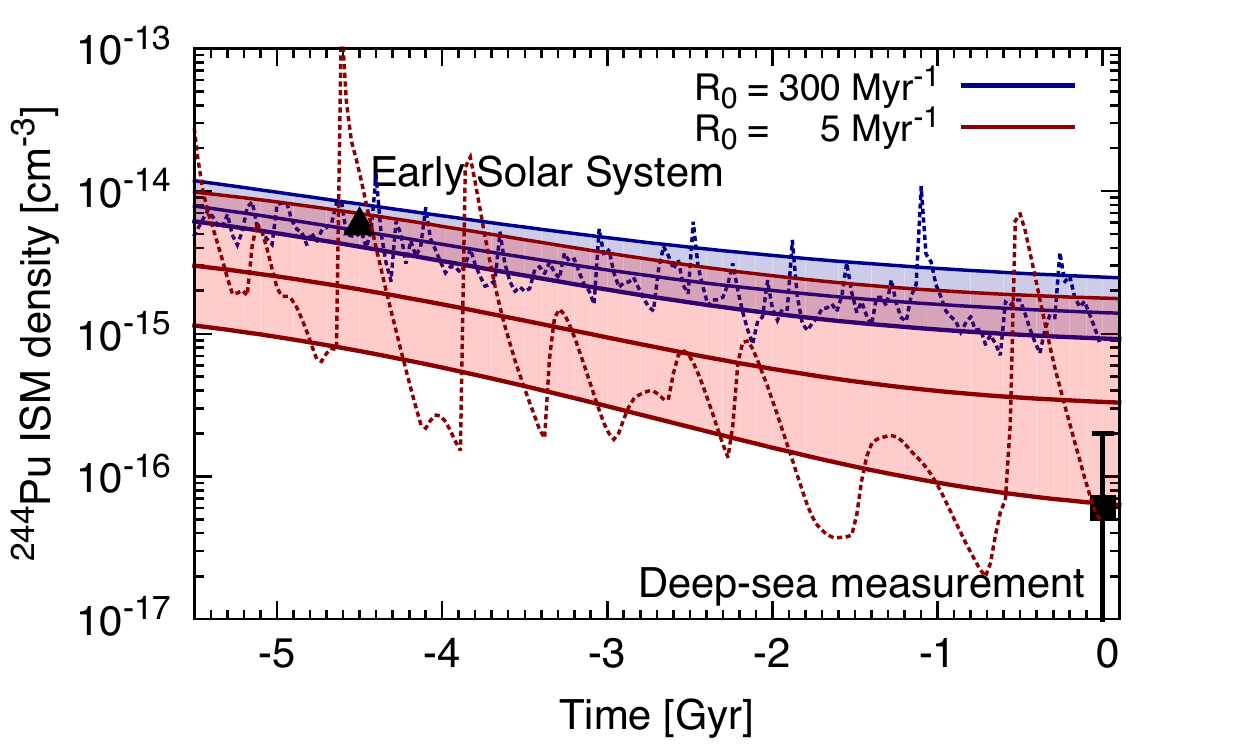}\\
\vspace{0.5cm}
\includegraphics[bb = 0 0 380 210, width=160mm]{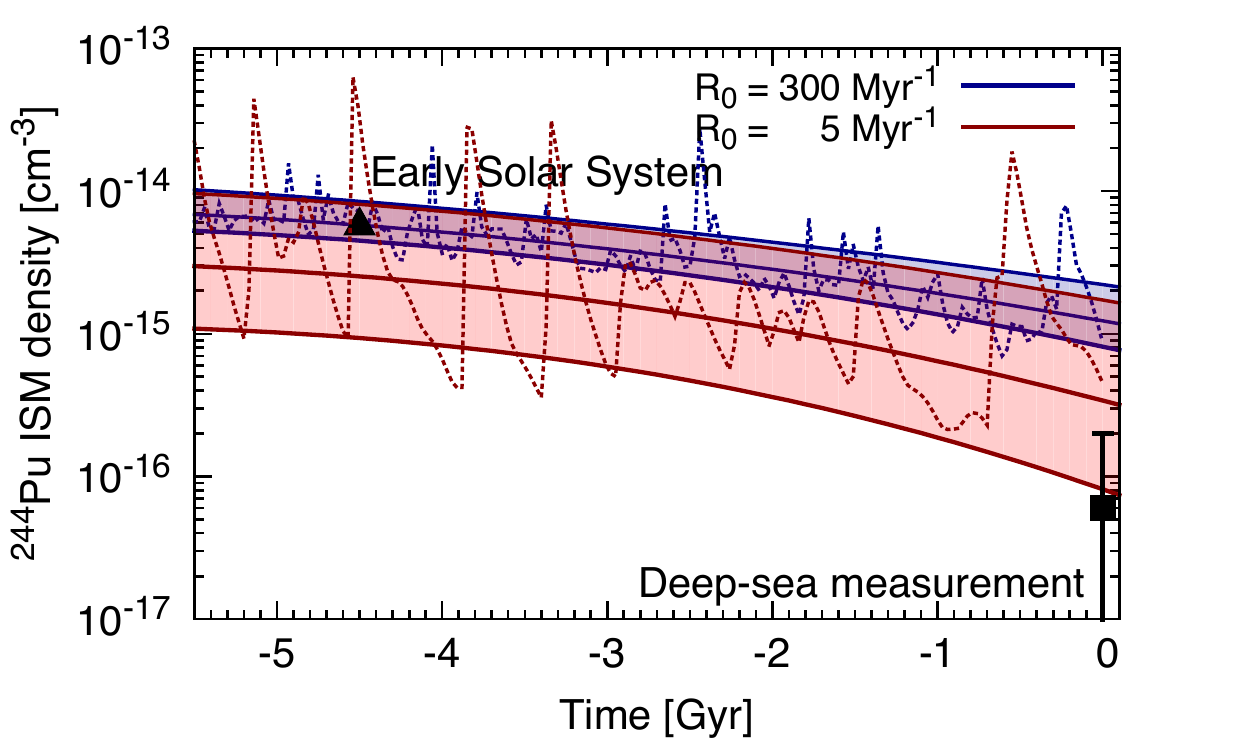}
\caption{
Same as Fig.~2 
but for the production rate following the SGRB rate~(top)
and the cosmic star formation history~(bottom).
}
\label{fig:rate1}
\end{figure*}

\begin{figure*}[h]
\includegraphics[bb = 0 0 380 210, width=160mm]{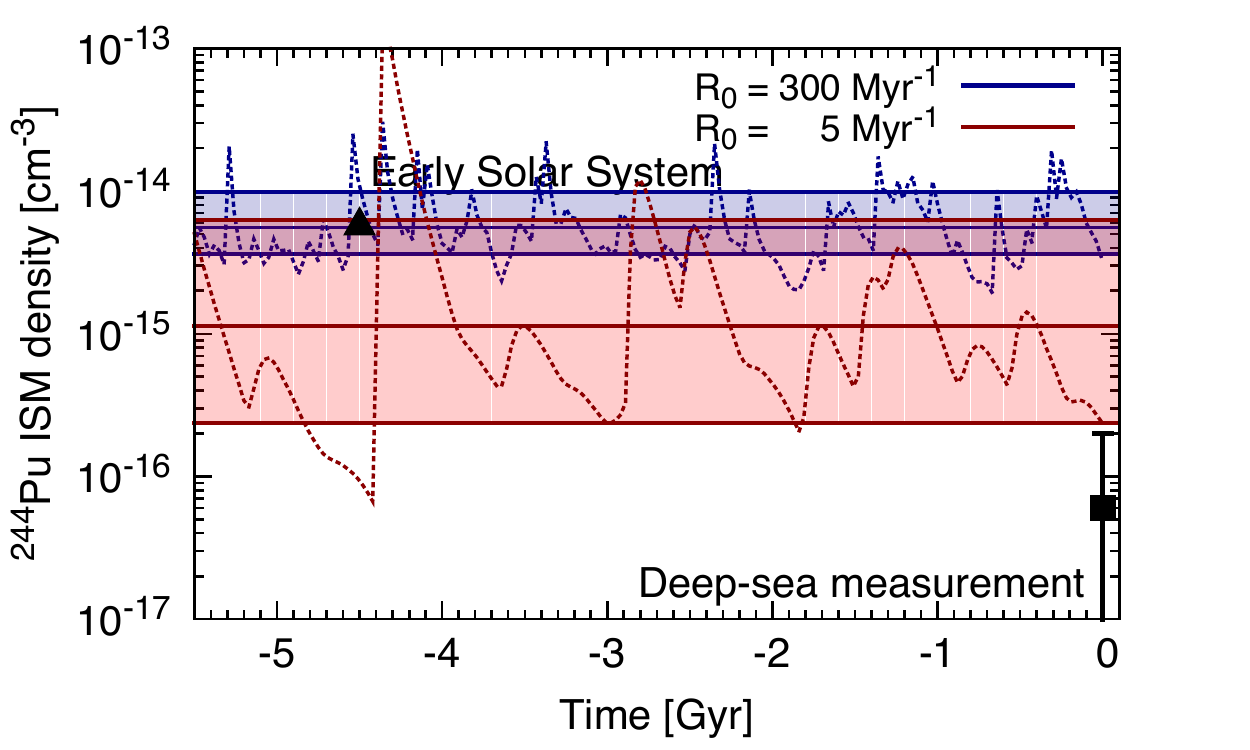}\\
\vspace{0.5cm}
\includegraphics[bb = 0 0 380 210, width=160mm]{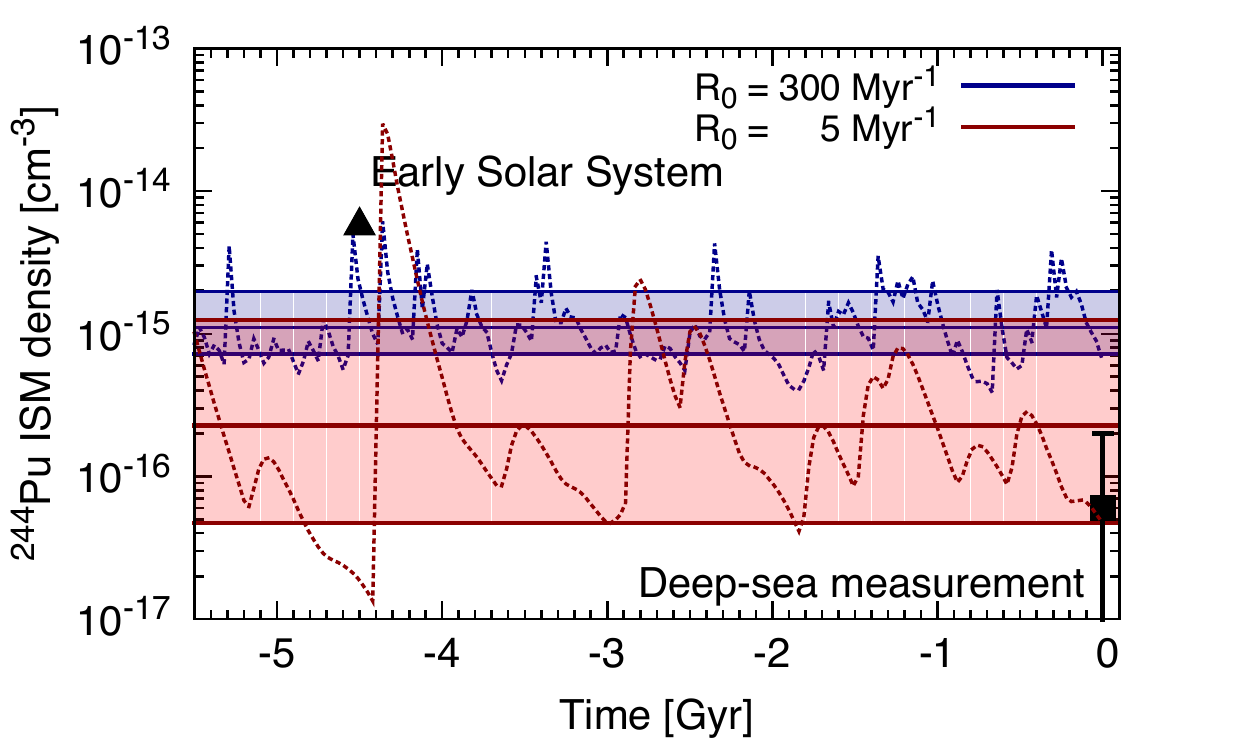}
\caption{
Same as Fig.~2
but for a constant production rate.
Here we assume $M_{{\rm ej},A\geq 90}=0.01M_{\odot}(R/10~{\rm Myr^{-1}})^{-1}$ for the top panel
and $M_{{\rm ej},A\geq 90}=0.05M_{\odot}(R/10~{\rm Myr^{-1}})^{-1}$ for the bottom panel.
}
\label{fig:rate2}
\end{figure*}

The time evolution of the mean abundance of $^{244}$Pu and its fluctuations
depend on the history of the production rate. Figure~\ref{fig:rate1}
depicts the rate-yield results with
the production rate following
the SGRB rate~(top panel) and the cosmic star formation
history~(bottom panel).
In both cases, the rate decreases with time for the epoch that we are interested in here.
The production rate at $4.6$~Gyr BP is higher than the current one by a factor
of $3$ for the former and $4$ for the latter.  As long as the event rate decreases
by these factors, the results do not depend sensitively on the choice of the
production-rate history.
However, if the event rate is constant with time,
very large fluctuations, i.e., small event rates, are required to be
consistent with both the ESS and current measurements as shown in
Fig.~\ref{fig:rate2}.

\section{Conclusion and Discussion}
We estimate the expected median value of the $^{244}$Pu abundances and fluctuations 
around this value in the high-rate/low-yield and low-rate/high-yield models 
and we show that while the current and ESS 
abundances are naturally explained within the low-rate/high-yield 
scenario they are incompatible with the high-rate/low-yield~(cc-SNe) model.
Remarkably, the rates and masses estimated here are fully consistent with astronomical 
observations of compact binary mergers.
In fact most of the overlap between the allowed $^{244}$Pu region and the overall $r$-process
production range is just in this part of the astrophysical parameter phase space describing
compact binary mergers and macronova ejection estimates. 
This result is independent of the choice of the efficiency and diffusion parameters.

Compact binary mergers, which we can conclude are the sources of heavy $r$-process nucleosynthesis,
are also the prime candidates of sources for the upcoming gravitational wave  detectors, advanced LIGO/Virgo and KAGRA.
Our estimates provide an upper limit to the expected detection rate (assuming a detection horizon distance of $200$~Mpc):
$R_{\rm GW}\leq 30~{\rm yr^{-1}}$.
The estimated ejected mass in each event is significant and this
implies that  macronovae~\citep{li1998ApJ, metzger2010MNRAS,barnes2013ApJ, tanaka2013ApJ}
and radio flares~\citep{nakar2011Nature}
associated with the gravitational-wave merger events will be detectable.

\vspace{1.0cm}

{\bf Acknowledgements}
We gratefully acknowledge useful discussions with M. Eichler, K. Kashiyama, H. Kimura, E. Nakar, A. Wallner,
and D. Wanderman.
This work was supported by the ISF I-Core center for excellence in Astrophysics, by CBSF
-ISF grant and by a grant from ISA.

\appendix

\section{Interpretation of the deep sea measurement}
\cite{wallner2015NatCo}
measured the number of live $^{244}$Pu
particles in deep-sea archives, which consist of two samples,
crust and sediment. Using the crust sample
the estimated flux of $^{244}$Pu particles on the Earth's orbit is
$f_{\oplus}^{\rm Pu}=250^{+590}_{-205}~{\rm cm^{-2}~Myr^{-1}}$  and it is
$3000^{+12000}_{-2850}~{\rm cm^{-2}~Myr^{-1}}$ with the sediment sample.
The upper and lower values are $2\sigma$ limits. The crust sample spans an
accumulation time of $25$~Myr and the sediment one spans $1.64$~Myr.
The two estimates are consistent with each other within the $2\sigma$ level.
Here, we use only the crust sample since it statistically dominates over
the sediment one and it spans a longer accumulation time. Even if one combines the
two data, our conclusion does not change significantly.
The estimated flux with the two samples combined is $363^{+580}_{-205}~{\rm cm^{-2}~Myr^{-1}}$ where again we take $2\sigma$ limits.
This  is only slightly larger than the one estimated using the crust sample alone.

The value of  $f_{\oplus}^{\rm Pu}$, which is the mean value over the last $25$~Myr,
is converted to the $^{244}$Pu number density in the ISM around the Solar System as
\begin{eqnarray}
n_{\rm ISM}^{\rm Pu}=
\frac{f_{\oplus}^{\rm Pu}}{\epsilon v_{\rm rel}},
\approx & 6\times 10^{-17}~{\rm cm^{-3}}\left(\frac{f_{\oplus}^{\rm Pu}}{250~{\rm cm^{-2}~Myr^{-1}}}\right)
\left(\frac{v_{\rm rel}}{26~{\rm km/s}}\right)^{-1}
\left(\frac{\epsilon}{0.05}\right)^{-1},
\end{eqnarray}
where
$v_{\rm rel}$ is the relative velocity between the Solar System and
the ISM.
The efficiency $\epsilon$ takes into account
(i) the penetration efficiency of the $^{244}$Pu flux from the ISM outside
the heliosphere to the Earth's orbit and
(ii) the difference between the mean ISM density around the solar circle
and the local ISM density where the Solar System has been
traveling during the last $25$~Myr.
\cite{wallner2015NatCo}
used $\epsilon = 0.03$--$0.09$.
Here we use $\epsilon=0.05$. In the following, we discuss how this  efficiency
is estimated.

The dust flux of the ISM inside  the heliosphere have been measured by the {\it Ulysses}, {\it Galileo},
{\it Cassini}, and {\it Helios} spacecrafts~(see \citealt{mann2010ARA}
for a review).
The measurements  yield  a lower limit  on the  gas-mass to dust-mass ratio
in the local ISM~\citep{frisch2013EP&S},
$\sim150$, which is roughly consistent with
astronomical estimates  based on  star-light extinction of  nearby stars.
In addition, the  dust flux in a mass range of $5\times 10^{-14}$ to $10^{-12}$~g measured by {\it Cassini}
around $1$~AU agrees with
the one measured around $3$~AU by {\it Ulysses} during the same period~\citep{landgraf2003JGRA, altobelli2003JGRA}.
Based on these facts, we assume that the $^{244}$Pu flux
on  Earth's orbit is the same as the one at the heliopause.
We also assume  that the $^{244}$Pu abundance
in the ISM dust grains is independent of the grain size.

As the ISM is highly inhomogeneous, one should take into account the
depletion of the local ISM  relative to the average galactic ISM density when evaluating the implied radioactive
nuclide's density. The Solar System is currently
traveling inside a small interstellar cloud~\citep{frisch2011ARA}
with a mean density of $0.1$--$0.3~{\rm cm^{-3}}$.
This cloud is within the Local Bubble that has a very low mean density of $0.005~{\rm cm^{-3}}$ and a
radius of $60$--$100$~pc.
Based on the solar motion and the size of the Local Bubble~\citep{frisch2006ASTRA},
the Solar System has been in a very low density region for the last $3$--$10$~Myr.
Before that time it had been outside the Local Bubble,
where the ISM density is in the range of $0.1$--$1~{\rm cm^{-3}}$.
Therefore the ISM density surrounding the Solar System averaged over
the last $25$~Myr can be estimated as $0.05$--$0.9~{\rm cm^{-3}}$.

Based on the above consideration, a plausible range of $\epsilon$ is $0.05$ to $0.9$.
Here we use $\epsilon=0.05$ as a reference value that corresponds to a conservative choice in terms of  the depletion of $^{244}$Pu in the ISM.
Clearly this is the largest source of uncertainty in our estimates.
Still  this uncertainty does not affect the qualitative nature of our results.
In Sec.~\ref{sec:sen} we also consider $\epsilon=0.01$. Even with this low value (see Fig.~\ref{fig:epsilon}), 
cc-SNe are only marginally consistent  with the deep-sea the measurements.
We note also that the penetration of ISM dust to the Earth's orbit in
deep-sea archives is confirmed by the observation of a spike of live $^{60}$Fe~(half-live of $2.6$~Myr)
in the deep-sea crust~\citep{knie2004PRL}
and sediments~(\citealt{fitoussi2008PRL}; Wallner et~al. in prep).
This  $^{60}$Fe
spike is interpreted as direct ejecta of a close-by supernova about $2.5$~Myr BP.


\begin{thebibliography}{45}
\expandafter\ifx\csname natexlab\endcsname\relax\def\natexlab#1{#1}\fi

\bibitem[{{Abadie} {et~al.}(2012){Abadie}, {Abbott}, {Abbott}, {Abbott},
  {Abernathy}, {Accadia}, {Acernese}, {Adams}, {Adhikari}, {Affeldt}, \&
  et~al.}]{abadie2012PRD}
{Abadie}, J., {et~al.} 2012, \prd, 85, 082002

\bibitem[{{Altobelli} {et~al.}(2003){Altobelli}, {Kempf}, {Landgraf}, {Srama},
  {Dikarev}, {Kr{\"u}ger}, {Moragas-Klostermeyer}, \&
  {Gr{\"u}n}}]{altobelli2003JGRA}
{Altobelli}, N., {Kempf}, S., {Landgraf}, M., {Srama}, R., {Dikarev}, V.,
  {Kr{\"u}ger}, H., {Moragas-Klostermeyer}, G., \& {Gr{\"u}n}, E. 2003, Journal
  of Geophysical Research (Space Physics), 108, 8032

\bibitem[{{Arnould} {et~al.}(2007){Arnould}, {Goriely}, \&
  {Takahashi}}]{arnould2007PhR}
{Arnould}, M., {Goriely}, S., \& {Takahashi}, K. 2007, \physrep, 450, 97

\bibitem[{{Barnes} \& {Kasen}(2013)}]{barnes2013ApJ}
{Barnes}, J., \& {Kasen}, D. 2013, \apj, 775, 18

\bibitem[{{Bauswein} {et~al.}(2013){Bauswein}, {Goriely}, \&
  {Janka}}]{bauswein2013ApJa}
{Bauswein}, A., {Goriely}, S., \& {Janka}, H.-T. 2013, \apj, 773, 78

\bibitem[{{Berger} {et~al.}(2013){Berger}, {Fong}, \&
  {Chornock}}]{berger2013ApJ}
{Berger}, E., {Fong}, W., \& {Chornock}, R. 2013, \apjl, 774, L23

\bibitem[{{Burbidge} {et~al.}(1957){Burbidge}, {Burbidge}, {Fowler}, \&
  {Hoyle}}]{burbidge1957RvMP}
{Burbidge}, E.~M., {Burbidge}, G.~R., {Fowler}, W.~A., \& {Hoyle}, F. 1957,
  Reviews of Modern Physics, 29, 547

\bibitem[{{Cioffi} {et~al.}(1988){Cioffi}, {McKee}, \&
  {Bertschinger}}]{cioffi1988ApJ}
{Cioffi}, D.~F., {McKee}, C.~F., \& {Bertschinger}, E. 1988, \apj, 334, 252

\bibitem[{{Cowan} {et~al.}(1987){Cowan}, {Thielemann}, \&
  {Truran}}]{cowan1987ApJ}
{Cowan}, J.~J., {Thielemann}, F.-K., \& {Truran}, J.~W. 1987, \apj, 323, 543

\bibitem[{{Cowan} {et~al.}(1991){Cowan}, {Thielemann}, \&
  {Truran}}]{cowan1991PhR}
---. 1991, \physrep, 208, 267

\bibitem[{{Eichler} {et~al.}(1989){Eichler}, {Livio}, {Piran}, \&
  {Schramm}}]{eichler1989Nature}
{Eichler}, D., {Livio}, M., {Piran}, T., \& {Schramm}, D.~N. 1989, \nat, 340,
  126

\bibitem[{{Eichler} {et~al.}(2015){Eichler}, {Arcones}, {Kelic}, {Korobkin},
  {Langanke}, {Marketin}, {Martinez-Pinedo}, {Panov}, {Rauscher}, {Rosswog},
  {Winteler}, {Zinner}, \& {Thielemann}}]{eichler2015ApJ}
{Eichler}, M., {et~al.} 2015, \apj, 808, 30

\bibitem[{{Fitoussi} {et~al.}(2008){Fitoussi}, {Raisbeck}, {Knie},
  {Korschinek}, {Faestermann}, {Goriely}, {Lunney}, {Poutivtsev}, {Rugel},
  {Waelbroeck}, \& {Wallner}}]{fitoussi2008PRL}
{Fitoussi}, C., {et~al.} 2008, Physical Review Letters, 101, 121101

\bibitem[{{Freiburghaus} {et~al.}(1999){Freiburghaus}, {Rosswog}, \&
  {Thielemann}}]{freiburghaus1999ApJ}
{Freiburghaus}, C., {Rosswog}, S., \& {Thielemann}, F.-K. 1999, \apjl, 525,
  L121

\bibitem[{{Frisch} {et~al.}(2011){Frisch}, {Redfield}, \&
  {Slavin}}]{frisch2011ARA}
{Frisch}, P.~C., {Redfield}, S., \& {Slavin}, J.~D. 2011, \araa, 49, 237

\bibitem[{{Frisch} \& {Slavin}(2006)}]{frisch2006ASTRA}
{Frisch}, P.~C., \& {Slavin}, J.~D. 2006, Astrophysics and Space Sciences
  Transactions, 2, 53

\bibitem[{{Frisch} \& {Slavin}(2013)}]{frisch2013EP&S}
---. 2013, Earth, Planets, and Space, 65, 175

\bibitem[{{Goriely}(1999)}]{goriely1999A&A}
{Goriely}, S. 1999, \aap, 342, 881

\bibitem[{{Hopkins} \& {Beacom}(2006)}]{hopkins2006ApJ}
{Hopkins}, A.~M., \& {Beacom}, J.~F. 2006, \apj, 651, 142

\bibitem[{{Hotokezaka} {et~al.}(2013){Hotokezaka}, {Kiuchi}, {Kyutoku},
  {Okawa}, {Sekiguchi}, {Shibata}, \& {Taniguchi}}]{hotokezaka2013PRDa}
{Hotokezaka}, K., {Kiuchi}, K., {Kyutoku}, K., {Okawa}, H., {Sekiguchi}, Y.-i.,
  {Shibata}, M., \& {Taniguchi}, K. 2013, \prd, 87, 024001

\bibitem[{{Kim} {et~al.}(2015){Kim}, {Perera}, \& {McLaughlin}}]{kim2015MNRAS}
{Kim}, C., {Perera}, B.~B.~P., \& {McLaughlin}, M.~A. 2015, \mnras, 448, 928

\bibitem[{{Knie} {et~al.}(2004){Knie}, {Korschinek}, {Faestermann}, {Dorfi},
  {Rugel}, \& {Wallner}}]{knie2004PRL}
{Knie}, K., {Korschinek}, G., {Faestermann}, T., {Dorfi}, E.~A., {Rugel}, G.,
  \& {Wallner}, A. 2004, Physical Review Letters, 93, 171103

\bibitem[{{Landgraf} {et~al.}(2003){Landgraf}, {Kr{\"u}ger}, {Altobelli}, \&
  {Gr{\"u}n}}]{landgraf2003JGRA}
{Landgraf}, M., {Kr{\"u}ger}, H., {Altobelli}, N., \& {Gr{\"u}n}, E. 2003,
  Journal of Geophysical Research (Space Physics), 108, 8030

\bibitem[{{Lattimer} \& {Schramm}(1974)}]{lattimer1974ApJ}
{Lattimer}, J.~M., \& {Schramm}, D.~N. 1974, \apjl, 192, L145

\bibitem[{{Li} \& {Paczy{\'n}ski}(1998)}]{li1998ApJ}
{Li}, L.-X., \& {Paczy{\'n}ski}, B. 1998, \apjl, 507, L59

\bibitem[{{Li} {et~al.}(2011){Li}, {Chornock}, {Leaman}, {Filippenko},
  {Poznanski}, {Wang}, {Ganeshalingam}, \& {Mannucci}}]{li2011MNRAS}
{Li}, W., {Chornock}, R., {Leaman}, J., {Filippenko}, A.~V., {Poznanski}, D.,
  {Wang}, X., {Ganeshalingam}, M., \& {Mannucci}, F. 2011, \mnras, 412, 1473

\bibitem[{{Lodders} {et~al.}(2009){Lodders}, {Palme}, \& {Gail}}]{lodders2009}
{Lodders}, K., {Palme}, H., \& {Gail}, H.-P. 2009, Landolt B{\"o}rnstein, 44

\bibitem[{{Mann}(2010)}]{mann2010ARA}
{Mann}, I. 2010, \araa, 48, 173

\bibitem[{{McMillan}(2011)}]{mcMillan2011MNRAS}
{McMillan}, P.~J. 2011, \mnras, 414, 2446

\bibitem[{{Metzger} {et~al.}(2010){Metzger}, {Mart{\'{\i}}nez-Pinedo},
  {Darbha}, {Quataert}, {Arcones}, {Kasen}, {Thomas}, {Nugent}, {Panov}, \&
  {Zinner}}]{metzger2010MNRAS}
{Metzger}, B.~D., {et~al.} 2010, \mnras, 406, 2650

\bibitem[{{Nakar} \& {Piran}(2011)}]{nakar2011Nature}
{Nakar}, E., \& {Piran}, T. 2011, \nat, 478, 82

\bibitem[{{Pan} \& {Scannapieco}(2010)}]{pan2010ApJ}
{Pan}, L., \& {Scannapieco}, E. 2010, \apj, 721, 1765

\bibitem[{{Paul} {et~al.}(2001){Paul}, {Valenta}, {Ahmad}, {Berkovits},
  {Bordeanu}, {Ghelberg}, {Hashimoto}, {Hershkowitz}, {Jiang}, {Nakanishi}, \&
  {Sakamoto}}]{paul2001ApJ}
{Paul}, M., {et~al.} 2001, \apjl, 558, L133

\bibitem[{{Qian} \& {Wasserburg}(2007)}]{qian2007PhR}
{Qian}, Y.-Z., \& {Wasserburg}, G.~J. 2007, \physrep, 442, 237

\bibitem[{{Rosswog}(2013)}]{rosswog2013RSPTA}
{Rosswog}, S. 2013, Royal Society of London Philosophical Transactions Series
  A, 371, 20272

\bibitem[{{Roy} \& {Kunth}(1995)}]{roy1995A&A}
{Roy}, J.-R., \& {Kunth}, D. 1995, \aap, 294, 432

\bibitem[{{Scalo} \& {Elmegreen}(2004)}]{scalo2004ARA}
{Scalo}, J., \& {Elmegreen}, B.~G. 2004, \araa, 42, 275

\bibitem[{{Sneden} {et~al.}(2008){Sneden}, {Cowan}, \&
  {Gallino}}]{sneden2008ARA}
{Sneden}, C., {Cowan}, J.~J., \& {Gallino}, R. 2008, \araa, 46, 241

\bibitem[{{Tanaka} \& {Hotokezaka}(2013)}]{tanaka2013ApJ}
{Tanaka}, M., \& {Hotokezaka}, K. 2013, \apj, 775, 113

\bibitem[{{Tanvir} {et~al.}(2013){Tanvir}, {Levan}, {Fruchter}, {Hjorth},
  {Hounsell}, {Wiersema}, \& {Tunnicliffe}}]{tanvir2013Nature}
{Tanvir}, N.~R., {Levan}, A.~J., {Fruchter}, A.~S., {Hjorth}, J., {Hounsell},
  R.~A., {Wiersema}, K., \& {Tunnicliffe}, R.~L. 2013, \nat, 500, 547

\bibitem[{{Turner} {et~al.}(2007){Turner}, {Busfield}, {Crowther}, {Harrison},
  {Mojzsis}, \& {Gilmour}}]{turner2007E&PSL}
{Turner}, G., {Busfield}, A., {Crowther}, S.~A., {Harrison}, M., {Mojzsis},
  S.~J., \& {Gilmour}, J. 2007, Earth and Planetary Science Letters, 261, 491

\bibitem[{{Wallner} {et~al.}(2015){Wallner}, {Faestermann}, {Feige},
  {Feldstein}, {Knie}, {Korschinek}, {Kutschera}, {Ofan}, {Paul}, {Quinto},
  {Rugel}, \& {Steier}}]{wallner2015NatCo}
{Wallner}, A., {et~al.} 2015, Nature Communications, 6, 5956

\bibitem[{{Wanderman} \& {Piran}(2015)}]{wanderman2015MNRAS}
{Wanderman}, D., \& {Piran}, T. 2015, \mnras, 448, 3026

\bibitem[{{Yang} {et~al.}(2015){Yang}, {Jin}, {Li}, {Covino}, {Zheng},
  {Hotokezaka}, {Fan}, {Piran}, \& {Wei}}]{yang2015NatCo}
{Yang}, B., {et~al.} 2015, Nature Communications, 6, 7323

\bibitem[{{Yang} \& {Krumholz}(2012)}]{yang2012ApJ}
{Yang}, C.-C., \& {Krumholz}, M. 2012, \apj, 758, 48

\end{thebibliography}
\end{document}